# Digital Twins in Graphene' Technology

Elena F. Sheka

https://orcid.org/0000-0003-0308-6260

**Abstract**
The Digital Twins concept in science has a long history that goes back to the beginnings of now widely accepted modelling. The ever-expanding amount of digital data accompanying modelling could not but cause a qualitative transition from modelling, subordinated to the goal of reproducing a real object, to an equal-right Digital Twins concept that offers an independent view of the real object. This chapter presents the first example of such a conceptual rethinking, using the example of material science of high-tech graphene materials.

## Introduction

Digital twins burst into our lives in 2002 from the podium of a Society of Manufacturing Engineers conference with the voice of Michael Greaves [1]. Although the term itself was proposed much earlier [2], it was Greaves who filled it with new content that presents the digital twins (DTs) today. Grieves proposed the DT as the conceptual model underlying product lifecycle management. Eight years later John Vickers suggested the DTs concept [3], which has been largely developed since. From that moment on, a wide march of this concept began in the industry, health, construction, business, education, social life, etc. This book will present the successes of this concept in various fields of human activity. In order not to duplicate other contributors to the book, I will limit myself to references to the most representative reviews [4-13], but a few.

Differing in nuances, a general presentation of the DTs concept concerns the trinity of physical object, virtual/digital object and the connection between them. The connections is provided by the data that flows from the physical object to the digital/virtual one and information that is available from the digital/virtual object to the physical environment. This concept, which is new for large massive fields of the human activity, has been widely exploited in academic researches since the first computer became available. Known as simulations or modelling, the concept implementation has provided a drastic development of the academic studies, related to the natural science, in particular.

Today it is impossible to imagine modern physics, chemistry (as well as they the same but with prefixes *bio*- and *geo*-) and material science without modelling. A huge leap in the development of computational programs and computing tools that has taken place over the past half century has led to the fact that many previously predominantly empirical sciences have become virtual-empirical, while some of them predominantly virtual. Despite such rapid development, the relationship between the real object and its model, established as a subordinate for the model, has not changed until recently. As it turned out, this circumstance significantly limits the further development of science in the case of its predominantly virtual nature. And here the DTs concept in the above wording enters the scene. Indeed, physical and virtual objects form the basis of both the modelling and the DTs concept. The difference between these two approaches is contained in a different sense, which is embedded in the understanding of the connection between these two objects. Using a grammatical term, it can be represented as complex (principal-subordinate) one in the first case and compound (equal-equal) in the second. In every language, substituting one sentence for another changes the meaning of the spoken speech. The same is true in the case of science. This chapter discusses precisely this change in semantic speech, which is introduced by the DTs concept into science, using the example of the virtual chemical physics of graphene [14].

Despite the widespread use of modelling in science, after a careful study of the available literature [12, 15, 16], I found only one reference concerning the DT concept [17], which precedes our works, however, in this case also related mainly not to polymer science, but to polymer reaction engineering. However, the pioneer of the DTs concept, Greaves, in one of his interviews given in 2018, said [1]: "It does not have to be an all-or-nothing project. There is a wide range of information that I can collect and process with the twin. Digital twins could also be used in very specific, very limited scenarios". Graphenics has occurred to be one of these scenarios. Graphene is an outstanding academic project, strongly oriented on high-tech technology [18]. Heralded in 2012 with a list of promising applications on 850 pages and receiving an exclusive for material science financial support, Graphene Flagship faced insurmountable difficulties in fulfilling its obligations. As a result, the project, first divided into bad and good graphene [19], and then practically completely abandoned good one, referring to the dishonesty of graphene material manufacturers [20], turned out to be practically unfulfilled by the end. And one of the reasons for this plight is the virtual nature of graphenics, oversaturated with simulation results, which have contributed to the development of this field in the wrong direction. The real reasons for the failure were revealed by removing the models from the subordinate state and giving them the value of DTs equal to real objects. This chapter describes the result of such a transformation, which was the first application of the DTs concept in the academic materials science.

## Graphene as a high-tech product

The name 'graphene' has a wide meaning covering a large set of different objects related to the modern graphenics [21]. The first concerns a honeycomb package of carbon atoms just forming bare graphene domains of different shape and size. One of such domain is shown in Figure 1.1a. Graphene crystal presents the next family member, which is simply a large domain, the linear dimensions of which exceed all significant size-limiting dimensions, starting from which such a domain can be described strictly in the language of solid-state physics [22]. In practice, empirical graphene crystal, obtained, say, by peeling away one layer of graphite with adhesive tape, is a one-atom thick sheet, linear size of which is usually over 1$\mu$. The domain edge atoms are valence not saturated thus possessing dangling bonds. In reality, these bonds are usually terminated by heteroatoms and chemical groups available in the object surrounding. In contrast, theoretical graphene crystal is a bare flat graphene domain generated by multiplication of an elementary cell consisting of two carbon atoms under the conditions of translational periodic invariance [23].

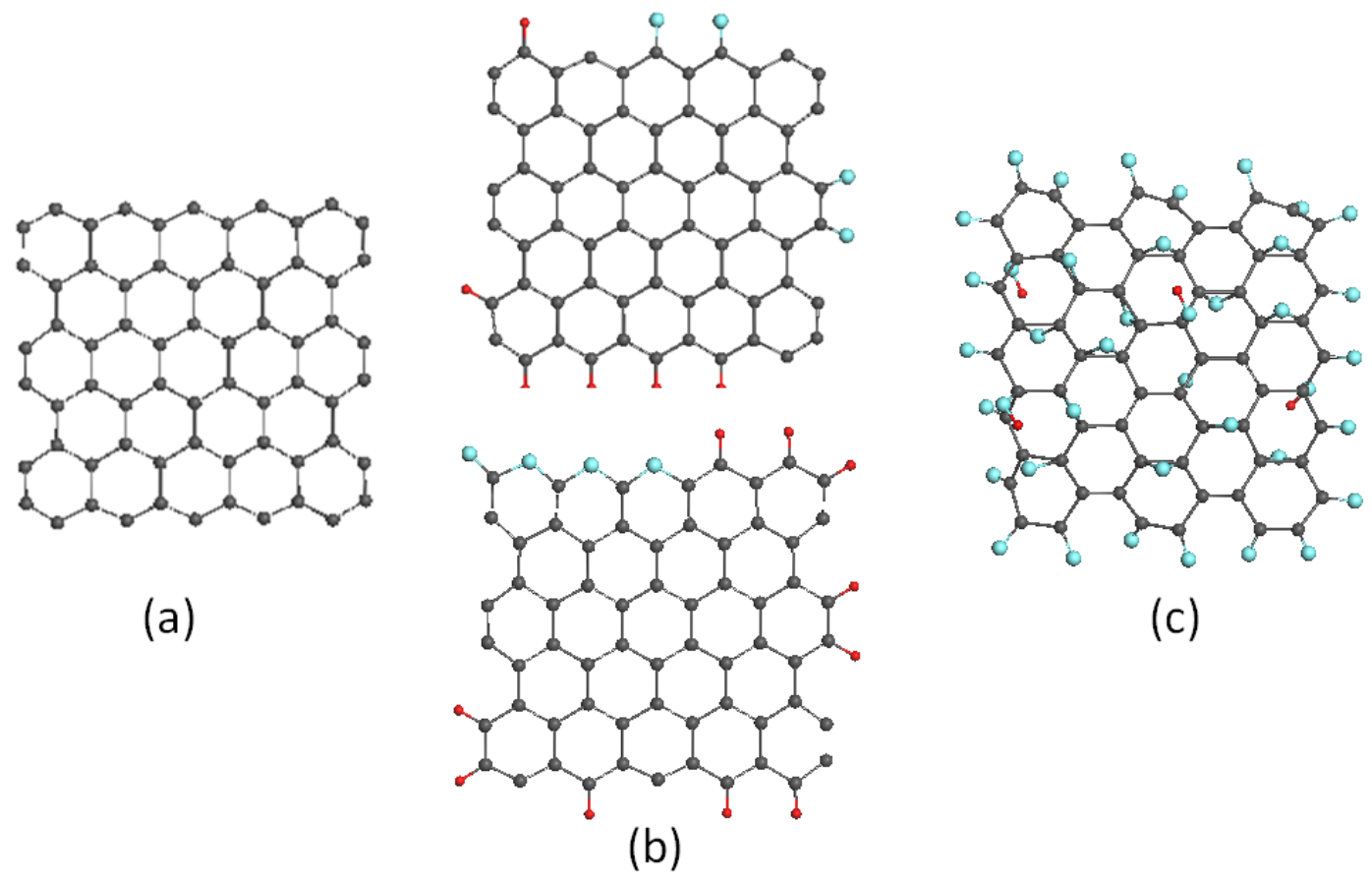

**Figure 1.1**. Molecular graphene: a. Bare graphene domain (5,5) NGr ($C_{66}$); b. Reduced graphene oxide on the example of basic structural units of shungite carbon ($C_{66}O_4H_6$ - top) and anthraxolite ($C_{66}O_4H_{10}$ - bottom); c. Basic structural unit of graphene oxide ($C_{66}O_{36}H_4$). Molecules' formular content is given in brackets. Grey, blue, and red balls mark carbon, oxygen, and hydrogen atoms.

The next family member is presented with graphene molecules. As previously, graphene domains lay the foundation of the species. However, the domains linear size is less than that of the above-mentioned size-characteristic parameters, which distinguishes the molecular and crystalline communities. The molecular part is extremely large involving all possible derivatives of graphene domains. Since the latter is a topochemical polytarget species [24], the number of polyderivatives is practically endless. Nevertheless, the latter can be divided into two subclasses, which are structurally distinguished, namely, the subclass of $sp^2$ necklaced graphene molecules, edge atoms of whose domains are involved in chemical reactions only [25] (Figure 1.1.b), and the subclass of $sp^3$ corpus graphene molecules, for which not only edge atoms, but also basal-plane atoms of the domains participate in the chemical modification [26, 27] (Figure 1.1.c). Besides this the most important graphene family members mentioned above, which are real products, there are graphene films variously deposited on different substrates [21, 23]. However, these films are objects for restricted lab-studies only and will not be considered in what follows.

The high practical appeal of using graphene in modern technology is based on the exceptional properties of its crystal [23] and molecules [14, 28]. Readers interested in this issue are referred to the above monographs. However, the practical implementation of this intention faced difficulties from the first steps. Thus, there was an understanding that a graphene molecule and a graphene crystal are completely different objects, which forced one to divide the relation to the graphene material into two parts that provide low performance (*lp*) and high performance (*hp*) technologies, spreading the implementation time of the corresponding technologies by decades [19]. The main faced-to problem concerned the production of graphene material. Thus, the adhesive-tape peeling of graphite can not be seriously considered as the source of graphene mass production for *hp* technology. On the other hand, the production of molecular graphene for *lp* technologies, the need of which was realized in physical laboratories mainly, was correlated with the presence of the initial product in the form of natural graphite, which needs splitting into nanosized sheets. The best way to achieve the problem, suggested with chemists, is the graphite oxidation [29] and hundreds of chemical laboratories around the world are engaged in the development of the most efficient and low-cost methods for the mass production of graphene oxide (GO) (see reviews [30-38] but a few). GO is a $sp^3$ corpus graphene species, while analyzing the molecule application, which has been rapidly developing all this time, it was found that not corpus GO, but necklaced reduced graphene oxide (rGO) was much more effective and interesting in many cases. The finding opened the road for the reduction of the previously synthesized GO transferring it to the wished rGO. And again, hundreds of chemical laboratories were engrossed in the development of efficient and low-cost graphene product known as rGO [39]. A full technological chain looks like the following:

*nanostructured Graphite* ↔ *Graphene oxide* ↔ *reduced Graphene oxide*

**Scheme 1**

Here sign ↔ means that the connected two processes are reversible. At the same time, once stimulated by the rapid development of graphenics, old $sp^2$ amorphous carbons experienced the writing of a new history. Thus, occurred that Scheme 1 was implemented in the Nature during the formation of shungite carbon deposits millions of years ago [40]. The same applies to coals and anthraxolites [41], carbon shells of silica [42] and other minerals, etc. At the same time, for the last century, the industry has managed to develop technologies for the tonnage production of synthetic $sp^2$ amorphous carbons, such as carbon black, activated carbon, carbon rubber fillers, etc [43]. It was found that rGOs of different necklace chemical content as well as graphene domain shape and size are the basis of all this carbon natural wealth [44]. Moreover, the humanity, without realizing it, is familiar with the body from the first bonfire, from the first shovel of coal thrown into the furnace, from the first filler of wheel tires, from the first ... and this listing can be continued indefinitely. The ash of a burnt fire, deposits of natural coal, synthetic black carbon, and, finally, a laboratory product of the chemical reduction of oxidized nanoscale graphite - this is a short list of a unique material, which is known as $sp^2$ amorphous carbon, the basic structural unit (BSU) of which is a polyatomic molecule, which should be attributed to rGO. Thus, returning to the beginning of history, the simplest method of obtaining a massive amount of rGO by burning food waste is being revived in the hands of materials scientists [45]. In contrast to the rGO, the GO does not exist in nature. The reason for this will be commented at the end of the chapter. For now, let's focus on the fact that chemical synthesis is the only source of this product.

In conclusion of this section, Figure 1.1 presents a visualization of the implementation of the processes according to Scheme 1 in the face of graphene structures models. A rightangle honeycomb sheet with five benzenoid units along armchair and zigzag edges consisting of 66 atoms ($C_{66}$) – the domain (5,5) NGr below - opens the collection in Figure 1.1a. The molecules $C_{66}O_4H_6$ and $C_{66}O_4H_{10}$ in Figure 1.1b are necklaced graphene molecules, designed on the derivatization of the (5,5) NGr concerning the edge carbon atoms only [25]. They present models of the BSU of shungite carbon and anthraxolite [41, 46], respectively, as well as are typical for numberless rGOs. The molecule $C_{66}O_{36}H_4$ in Figure 1.1c is a corpus graphene molecule designed as the result of derivatization of not only edge, but basal-plane atoms as well and presents one of possible presentations of numerous GOs [27].

Graphene molecules do not exist individually and, once of either natural or synthetic origin, are fully incorporated in carbon amorphous solids. The solids have a complicated multilevel structure [40, 41], which is schematically presented in panel A of Figure 1.2. The general picture presented in the left side of the panel is typical to $sp^2$ amorphous rGOs. As seen, individual BSUs are firstly stacked. Thus formed stacks agglomerate producing the solid body of highly variable porosity thus separating restricted regions of lower porosity (globules) in the total whole of the body. Three main structural characteristics of these solids are the linear size of their BSUs, $L_a$, the stack thickness, $L_c$, and the distance between BSU layers in stacks, $d$. Amorphous GO retains the general motive of rGO. *sp*$^3$ Corpus graphene molecules present its BSUs, $L_a$ of which is comparable with that of rGO. Agglomerating, these BSUs clearly reveal stack structure and various porosity of the solid, while differing in *d* more than twice [31, 47]. Presented on right part of panel A of the figure displays four-layer stack compositions of rGO and GO based on molecules $C_{66}O_4H_6$ and $C_{66}O_{36}H_4$ in Figure 1.1b and 1.1c, taken as the relevant BSUs. The interlayer distance *d* is minimal in both cases, constituting 0.34 nm and 0.72 nm, respectively, and corresponds to the touching of the van der Waals diameters of atoms located in adjacent layers. The figures are in a good agreement with experimental data for shungite carbon [41] and solid GO [48]. Attention is drawn to the difference in the images corresponding to the stack top views of rGO and GO, evidencing a completely different net configuration of the relevant covalent C-C bonds, *sp*$^2$ in rGO and *sp*$^3$ in GO.

A

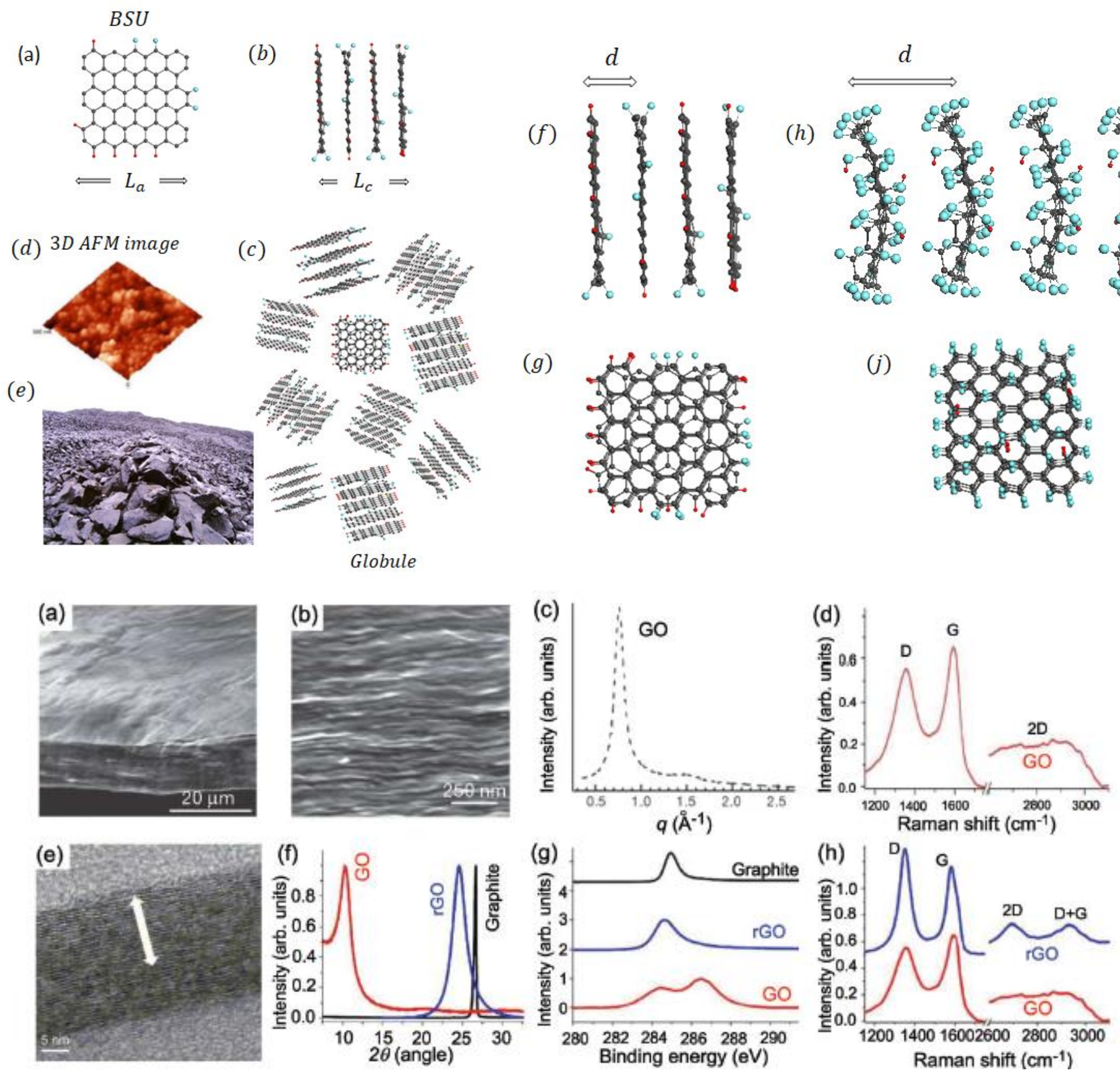

**Figure 1.2**. A. (a, b) Individual and four-layer stack of rGO BSU. (c). rGO stack globule. (d) General AFM image of shungite carbon. (e). Karelian shungite carbon deposit. (f, g) Side and top view of rGO stack, respectively. (h, j) Side and top view of Go stack, respectively.
B. (a) Low- and (b) high-resolution SEM side-view images of a 10 mm-thick GO paper. (c) X-ray diffraction pattern of the GO paper sample. (d) Raman spectrum of a typical GO paper. (e) Cross-section TEM images of a stack of rGO platelets. (f) Powder XRD patterns of graphite, GO, and rGO. (g) XPS characterization of rGO platelets. (h) Raman spectra of rGO (blue) and the GO reference sample (red). Reproduced with permission from ref. [31].

## The problem with graphene materials in practice

Reliable technological application of any material requires a mandatory test-on-individuality certificate of the latter. Particularly, in the case of the two amorphics that both are just black similar powders. Numerous in-depth laboratory analyses allowed filling this certificate with real content, the result of which is represented by the first row in panel B of Figure 1.2. [31, 35] For comparison, the corresponding data related to rGO are involved in the second row listing in the figure. As seen, Figures 1.2B(a), B(b), and B(c) clearly evidence layer structure of both solids, expectedly more waving in the GO case. Powder X-ray diffraction data on panel B(f) well support about twice difference in the interlayer distance of the solids, X-ray photoelectron spectroscopy on panel B(g) convincingly shows the change in chemical content of the solids caused by the oxygen contribution, comparable with carbon one, in the GO case. All these data taken together convincingly evidence that rGO and GO are completely different. However, when we are looking at Figure 1.2B(h), the moment of truth has come, since the observed identity of Raman spectra of the solids presented on it comes into sharp conflict with the above conclusion. As well known, Raman scattering spectroscopy is well sensitive to structure and chemical content of the object under study [49]. Accordingly, the observed similarity should be considered as an outstanding spectral event that requires a thorough consideration.

The situation is significantly complicated by the fact that a rather simple but greatly characteristic D-G-bands Raman doublet is largely exploited in graphenics as express-analysis justifying the belonging of the studied material to the family of graphene materials [50]. While not disputing that the GO belongs to graphene materials, we nevertheless do not consider the observed similarity of the spectra to be a mere coincidence. Moreover, this circumstance leads to a number of undesirable results concerning practical application of materials. A technologist, who is not able to distinguish rGO and GO by express Raman spectra analysis cannot guarantee the fulfillment of at least two mandatory requests: 1. The absence of a reversible transformation rGO $\leftrightarrow$ GO in due course of production and storage of the final product and 2. The absence of toxicity of the final product. This is vital in the use of rGO or GO in the manufacture of different medicals [51-53] and vaccines [54]. Suffice it to recall that rGO BSUs are stable, or sleeping, radicals [55], which can revive their activity at any moment under the action of surrounding, while GO is completely deactivated. It is also of great importance in the case of, for example, the use of carbon based hydrogels in the production of lithium batteries [56]. It is clear that rGO $\leftrightarrow$ GO processes greatly influence the efficiency of devices and their validity term.

To be able to fight and predict the possible negative consequences of the rGO and GO practical applications, we must answer the following question: “What do we still not know about rGO and GO, which would explain reasons for the identity of the Raman spectra of two substances, which are completely different both in structure and in chemical content?” As occurred, the DT concept allows getting the answer to this question [25, 57, 58].

## General grounds of the digital twins design

First used recently [59], the contradistinction of *compound* DT concept to *complex* molecular modelling has revealed a high efficiancy of the former to solve intricate complicated problems. It was naturally to apply to it looking for reasons of the identity of Raman spectra of rGO and GO and finding a way to discover this mystery. We look at the DT concept following Scheme 2

*Digital twins → Virtual device → IT product*.

Scheme 2

Here, DTs are molecular models under study, virtual device is a carrier of a selected software, IT product covers a large set of computational results related to the DTs under different actions in the light of the soft explored. The quality of IT product highly depends on how broadly and deeply the designed DTs cover all the knowledge concerning the object under consideration and how adequate is the virtual device to peculiarities of the object under study. The first requirement can be evidently met by a large set of the relevant models, in the current case up to a few hundreds, each consisting of $N$ atoms (up to 264). As for the virtual device, it should not contradict with the object nature and perform quantum-chemical calculations providing the establishing of equilibrium structure of the designed DTs and obtaining their spectra of IR absorption and Raman scattering related to $3N$ vibrational modes. Only semi-empirical programs based on the Hartree-Fock (HF) approximations and the theory of density functional (DFT) can cope with such a volume of cumbersome quantum-chemical calculations. However, the radical nature of most rGO and/or GO DTs, which turns them into open-shell electronic systems, forced to completely abandon DFT-based programs and pay attention only to programs based on the unrestricted Hartree-Fock (UHF) approximation. In the performed studies [25, 57, 58], the virtual device HF Specrodyn [60] was used.

**Spin-density algorithm of the Digital Twins design**

As told earlier, both rGOs and GOs are polyderivatives of bare honeycomb graphene domains. If suppose that all the domain atoms are equivalent (say, 66 atoms of the domain (5,5) NGr), the number of derivatives involving $n$ carbon atoms constitutes the number of combinations according to $n$ from the total number of atoms $N$. For, say, $n$=18 and $N$=66 this number is over million. And this situation is repeated at each sucessive reaction step over $n$. Guessing the desired structure from this huge number of isomorphs is impossible. The situation is saved by the fact that, in fact, the atoms of the domain are not equivalent, which is a manifestation of its radical nature, and involve $N_D$ effectively unpaired atoms, which are tightly connected with the total measure of spin density of the molecules [61]. UHF calculations make it possible to determine both the $N_D$ value as well of its fractures related to each carbon atoms, $N_{DA}$, thus supplying the atom with a highly important quntitive parameter. This parameter, named as the atomic chemical susseptibility (ACS), coincides with the free valency of the atom and determines its chemical activity. The value depends on the length of the relevant covalent C-C bond [62]. Concerning $sp^2$C-C bonds, ACS is nil until the bond length overcomes the critical value $R_{crit}$ = 1.395 Å. As occurred, usually UHF calculated bond lengths of the graphene domain fill the interval of 1.322-1.462 Å. The relative number of bonds, whose length exceeds $R_{crit}$ , constitutes 62% of the bond massive The resulting radicalization leads to a considerable amount of the effectively unpaired electrons of $N_D$ total number.

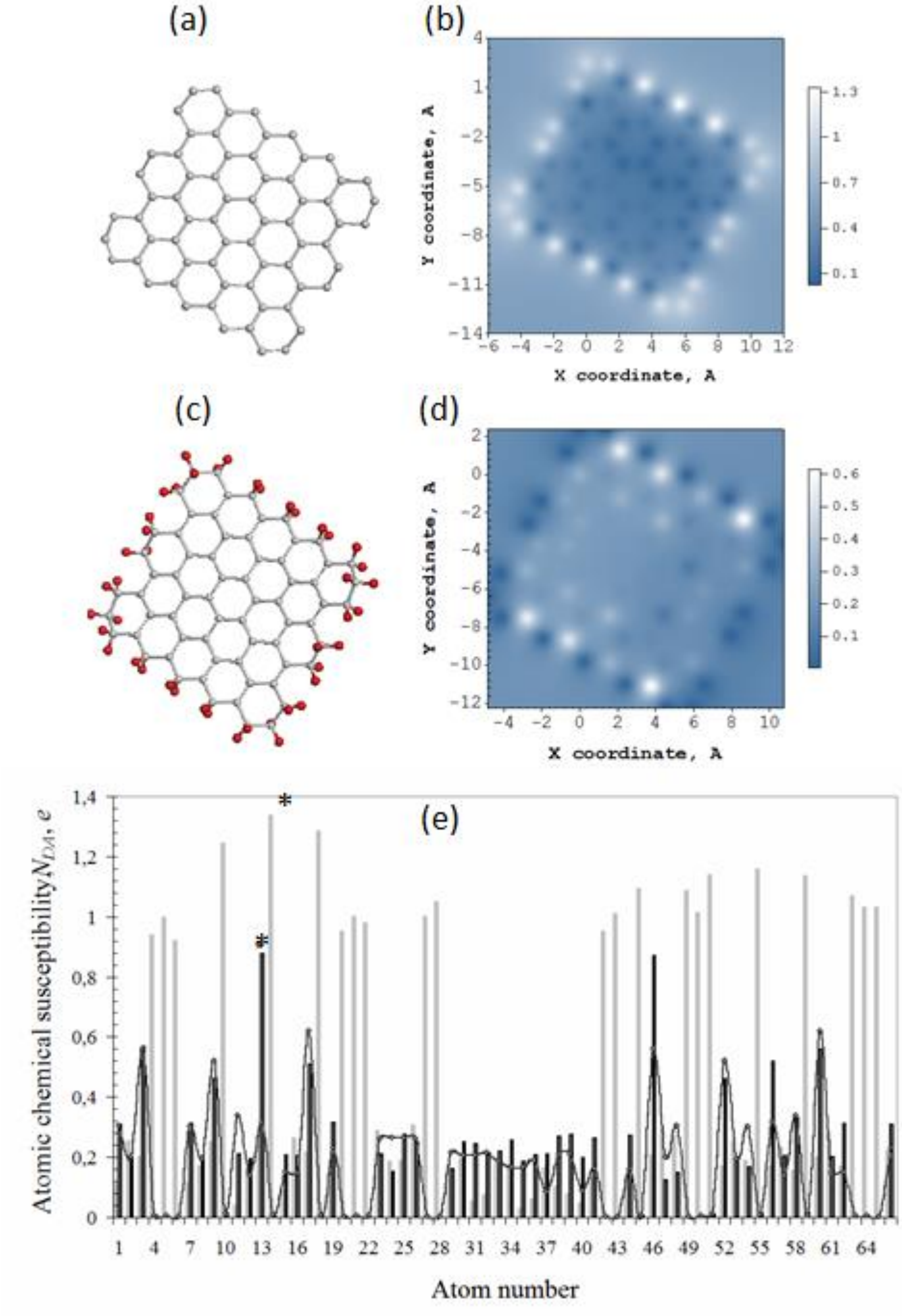

**Figure 1.3**. Equilibrium structures of bare domain (5, 5) NGr ($C_{66}$) (a), (5, 5) NGr, terminated by two hydrogen atoms per each edge one ($C_{66}H_{44}$) (c), and ACS $N_{DA}$ image maps over atoms in real space (b, d) as well as according to the atom number in the output files (e). Light grey histogram plots ACS data for $C_{66}$, while black one and curve with dots are related to $C_{66}H_{44}$. Scale bars match $N_{DA}$ values. UHF AM1 calculations.

Obviously, the atoms with the maximum $N_{DA}$ value will be the first target to enter any chemical reaction. Accordingly, an ACS spin-density algorithm was proposed [63] to match targets for successive reaction steps following high rank data in the output file. Schematicaly, the algorithm action is shown in Figure 1.3 on the example of the domain (5,5)NGr hydrogenation. Figure 1.3a exhibits a typical ACS image map presented by the $N_{DA}$ distribution over 66 atoms of the domain (5,5) NGr. The domain edges are not terminated and this map presents a typical chemical portrait of any bare graphene domain and exhibits the exceptional role of the circumference area, which looks like a typical ‘dangling bonds’ icon. At the same time, the ACS image map intensity in the basal plane is of ~0.3 *e* in average so that the basal plane should not be discounted when it comes to chemical modification of the domain. The absolute $N_{DA}$ values of the (5,5) NGr, shown by light grey plotting in Figure 1.3e, clearly exhibit 22 edge atoms involving 2x5 *zg* and 2x6 *ach* ones with the highest $N_{DA}$ thus marking the perimeter as the most active chemical space of the domain, successively terminated one by one firstly. The first step of the derivatization occurs on atom 14 (see star-marked light grey plotting in Figure 1.3e) according to $N_{DA}$ of the highest rank in the output file. The next step of the reaction involves the atom from the edge set as well, and this is continuing until either all the edge atoms are saturated or some of the basal ones come into play. In the case of the domain hydrogenation, all 44 steps are accompanied with the high-rank $N_{DA}$ list where edge atoms take the first place thus being terminated finally by hydrogen pairs [26]. Thus obtained hydrogen-necklaced (5,5) NGr molecule $C_{66}H_{44}$ is shown in Figure 1.3c alongside with the corresponding ACS image map in Figure 1.3d which reveals the transformation of brightly shining edge atoms in Figure 1.3b into dark spots. The addition of two hydrogen atoms to each of the edge ones saturates the valence of the latter completely, which results in zeroing $N_{DA}$ values as is clearly seen in Figure 1.3e. Thus, the chemical activity is shifted to the neighbouring basal-plane atoms. The curve in the figure exhibits free valence distribution over the molecule atoms in basal plane (see details in [61]). As seen in the figure, the hydrogenation will proceed on atom 13 (see star-marked black plotting). Evidently, the order in which carbon atoms come into per-step play depends on the chemical reagent, the fixing conditions for the graphene domain, as well as on *up*-and/or-*down* accessibility of the basal plane atoms. These issues are discussed in details with respect to the hydrogenation and oxidation of the domain (5,5) NGr [27, 28].

Once derivatized, both edge and basal-plane $sp^2$ carbon atoms of the pristine domains, which are valence undersaturated, become fully valence saturated, thus undergoing a transition to $sp^3$ hybridization. This results in a severe transformation of the plreviously flat benzenoid structure thus transfering the latter to a riffled cyclohexanoid one, which is clearly seen on the top views of rGO and GO in Figure 1.2A(g) and A(j). Having a general understanding of the mechanism of derivatization of graphene domains and remaining with the assumption that derivatization is a sequence of steps, the choice of targets in each of which is subject to the spin-density algorithm, we get a unique opportunity for the design of DTs, aimed at obtaining answers to well-defined questions. It is precisely this approach, recently applied to the rGO and GO, which made it possible to reveal the secret of the identity of the Raman spectra of these solids [57, 58]. Let us consider a step-by-step solution to this problem from the viewpoint of the DT concept.

## Vibrational spectra of rGOs and GOs in light of the Digital Twins concept

Exclusiveness of the situation with Raman spectra of rGO and GO causes a few questions to be answered, namely:

- Is the spectra identity a chemical-content effect?
- Is the spectra identity a structural effect?
- Is there any other reason for the effect?

It is clear that the answers to these questions affect the deep essence of both oxides, each of which is a broad classe of substances, and that they cannot be obtained empirically. In contrast, it is obvious that a wide chemical and structural modification of DTs makes it possible to find answers to the first two questions, after which it will be possible either to cancel the third one or to look for new ways of answering it. Let us consider how it was done in the considered case. To simplify further discussions, we classify rGO and GO as either polychrome or monochrome objects. The polychromicity concerns the variety of heteroatom content beyond the carbon one, which is typical to real products. The monochromicity will be used to limit the products heteroatomic content to mono-atomic one or to special chemical groups. So, in accordance with Scheme 2, DTs represent a wide range of models of both the individual BSUs of the rGO and GO and their stacks of different composition and different sizes. The digital device is a quantum-chemical software that implements the Hartree-Fock approximation in both versions concerning the unrestricted (UHF) and restricted (RHF) approximations. The IT product is a fully optimized DT structure on the one hand, and virtual one-phonon spectra of IR absorption and Raman scattering, on the other.

***rGO in light of the Digital Twins concept***

Two sets of designed DTs were considered. The first involves rGO DTs, based on the same graphene domain (5,5) NGr but differing with the heteroatom necklace content. The necklaces are both monochrome, consisting of either hydrogen or oxygen atoms, as well as of either hydroxylic or carboxylic units only, and polychrome when the content is mixed. Figure 1.4 presents a collection of IR and Raman spectra characteristic for the set. Through over the chapter, the virtual spectra are presented by stick-bars convoluted with Gaussian bandwidth of 10 $cm^{-1}$. Intensities are reported in arbitrary units, normalized per maximum values within each spectrum. Since the number of vibrational modes composing the spectra under consideration is too large, the excessive fine structure, statistically suppressed in practice, is covered by trend lines averaged over 50 next steps of $0.003 \div 0.010$ $cm^{-1}$ each. The figure is placed over a background composed of the relevant equilibrium DTs, the detailed study of which is discussed elsewhere [25, 57]. A comparative analysis of calculated data obtained completes the digital cycle. As a result, the following conclusions important for the DT analytics of rGO were obtained.

- The carbon atoms of graphene domains of the rGOs and heteroatoms of their necklaces contribute to the optical vibrational spectra quite differently: the vibrations of the former are mainly responsible for Raman spectra, while the IR spectra are assigned to the vibrations of the necklaces heteroatoms.
- Virtual IR spectra in Figure 1.4a present well distinguished spectral signatures of the monochrome rGOs related to the contribution of atomic hydrogens and oxygens as well as hydroxyls and carboxyls to the relevant rGOs necklaces.
- Virtual IR spectra of polychrome rGOs retain superpositional character of spectral signatures, although not providing a quantitative analysis of the necklace content, which is connected with the domain-stimulated influence of heteroatom additives on each other.
- Virtual results applied to empirical IR spectra of $sp^2$ amorphous carbons revealed that the necklace composition of the corresponding BSUs consists mainly of hydrogen and oxygen atoms. The variety of possible compositions of hydrogen- and oxygen-containing units, mainly of the latter, determines the wide variability of the empirical IR spectra [46].
- The DTs consist of almost a hundred of atoms, due to which eigenvectors of harmonic vibrations of the molecules involve many hundreds of components. This makes the attribution of the considered vibrational modes to individual chemical bonds, which is typical for molecular spectroscopy, inadequate. However, both virtual and empirical IR spectra of $sp^2$ amorphous carbons retain a certain similarity to the spectra of small molecules, which allows speaking about the selection of particular atomic compositions when generating absorption spectra. This feature allowed suggesting general frequencies kits of nanosize rGO instead of general frequencies of small molecules [64], implying the excitation of a set of bonds. The tabulated GFKs lay the foundation of express spectra analysis of rGO IR spectra.

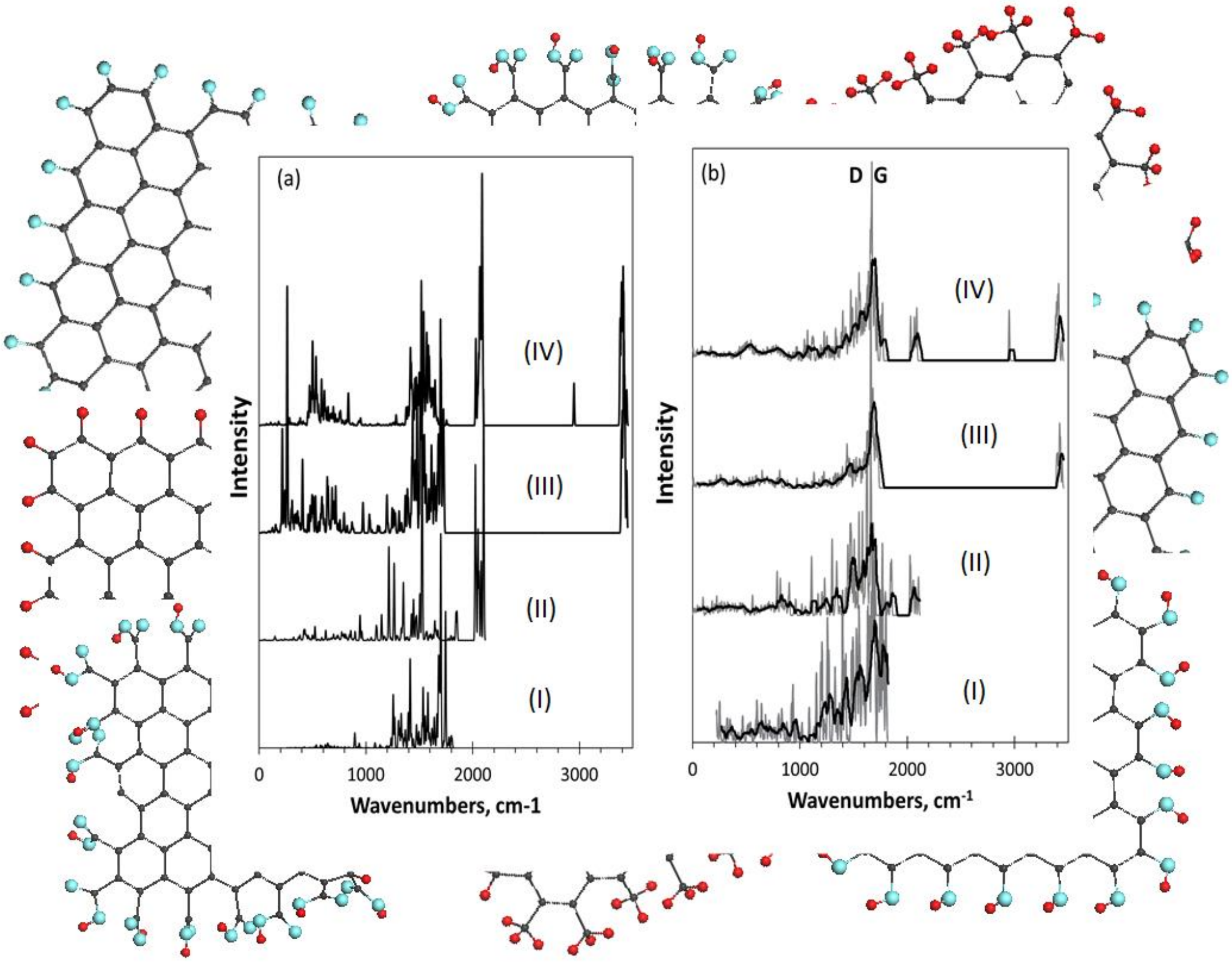

**Figure 1.4.** Virtual IR absorption (a) and Raman scattering (b) spectra of necklaced graphene molecules based on domain (5,5) NGr, presenting graphene hydride $C_{66}H_{22}$ (I), and three reduced graphene oxides $C_{66}O_{22}$ (II), $C_{66}(OH)_{22}$ (III), and $C_{66}(COOH)_{18}$.(IV). UHF AM1 calculations.

- Raman spectra of the considered DTs, part of which is presented in Figure 1.4b, in contrast to IR spectra, are more similar than different. The similarity is evidently provided with the same graphene domain (5,5) NGr of the studied DTs while the difference is explained by the different disturbance of the domain structure caused by the corresponding necklaces [25, 57]. The most important sign of these Raman spectra is the absence of characteristic D-G-doublet structure that is typical for rGOs in reality, as shown in Figure 1.2B(h). Thus, the Raman spectra of monolayer rGOs is not of a standard doublet image but is presented with a broad multiband structure.

The expansion of the set of considered DTs was a response to this surprise. It was natural to turn to the next element of the structure of amorphous rGOs, which concerns its stack level. The previous DTs set described above was added with two-layer and three-layer combinations of necklaced molecules based on the former domain (5,5) NGr, as well as individual molecules based on the enlarged domains (9,9) NGr and (11,11) NGr. The main results obtained for this group of DTs are summarized in Figure 1.5. As previously, the figure is placed over the collection of the corresponding DTs in use.

Raman spectrum in Figure 1.5a belongs to a DT designed with respect to synthetic rGO produced in due course of thermal explosion (TE-rGO) [46, 65]. The latter differs from the reference (5,5) NGr (see spectrum in Figure 1.5b) with not only the presence of complicated necklace, but with twice bigger size of the domain. As seen in the figure, this DT Raman spectrum drastically changes presenting the transformation of a broad multiband structure of the reference DT towards a single-band one. Such a transformation of Raman spectra is repeated for DTs involving bare and hydrogenated domains (9,9) NGr as well and becomes more pronounced in the case of bare domain (11,11) NGr [25]. The matter is connected with a typical size-effect concerning the Raman spectra of substances based on covalent bonds when the substance nature is transformed from molecularly amorphic to crystalline one [66, 67]. Actually, when the linear dimension of the domains is achieving and/or exceeds $L_{ph}$~15 nm, which is a free path of the high-frequency optical phonon G of the graphene crystal [68], the domain dynamics acquires the characteristic features of crystalline behavior and Raman spectra tend to adopt a single-band appearance in favor of G band corresponding to asymmetrical $sp^2$C-C stretchings, the limit case of which is presented with the narrow G phonon band in graphene crystal.

In contrast to one-layer molecules, two-layer hydrogenated (5,5) NGr is characterized with clearly seen generation of D-G-band doublet in Figure 1.5c. This change in the spectrum shape becomes more pronounced when the two-layer DT is replaced with a three-layer one. Therefore, the performed DT analysis has revealed that the characteristic standard D-G-band-doublet structure of the Raman spectra of rGOs is a consequence of its stacking structure. A thorough analysis has shown [25, 57] that the feature results from one more unique characteristics of carbon. It concernes the fact that the distance between adjacent layers in graphite-like structures is equal to the van der Waals diameter of a carbon atom and constitutes ≈0.335 nm. The distance is much larger than a standard size of a single $sp^3$C-C covalent bond of 0.153 nm, so that unpaired electrons of carbon atoms in the graphene domain basal planes of neighboring layers cannot be coupled while atoms themselves on their periphery contact each other. At the same time, vibrations of carbon atoms are not limited in space, due to which the out-of-plane displacements towards each other, which are supported with the occurrence of special out-of-plane phonons in a multilayer graphene crystal [69-72] stimulate the appearance of *dynamic* $sp^3$C-C bonds between the domain layers [25]. The corresponding out-of-plane vibrations lead to the emergence of a broad band in the spectrum of the stack of nanosized graphene domains, which is on the place of a quite narrow D band of multilayer graphene crystal [69]. Therefore, the doublet of the D–G bands of $sp^2$ amorphous carbons is a sign marking the stack structure of the substance. Evidently, the intensity of D band will increase when thickness of stacks increases, which was confirmed empirically in the case of anthraxolites [73].

Thus, answering the questions posed at the beginning of the section, we can state the following. The standard form of Raman spectra of rGOs has a structural origin and is a consequence of the stack structure of solid. The D and G bands are due to stretching vibrations of $sp^3$C-C and $sp^2$C-C bonds, respectively. In this case, $sp^2$ bonds constitute the main structural array of the rGO graphene domains, while $sp^3$ bonds are stimulated with out-of-plane displacements of the adjacent layer atoms towards each other and have a dynamic character.

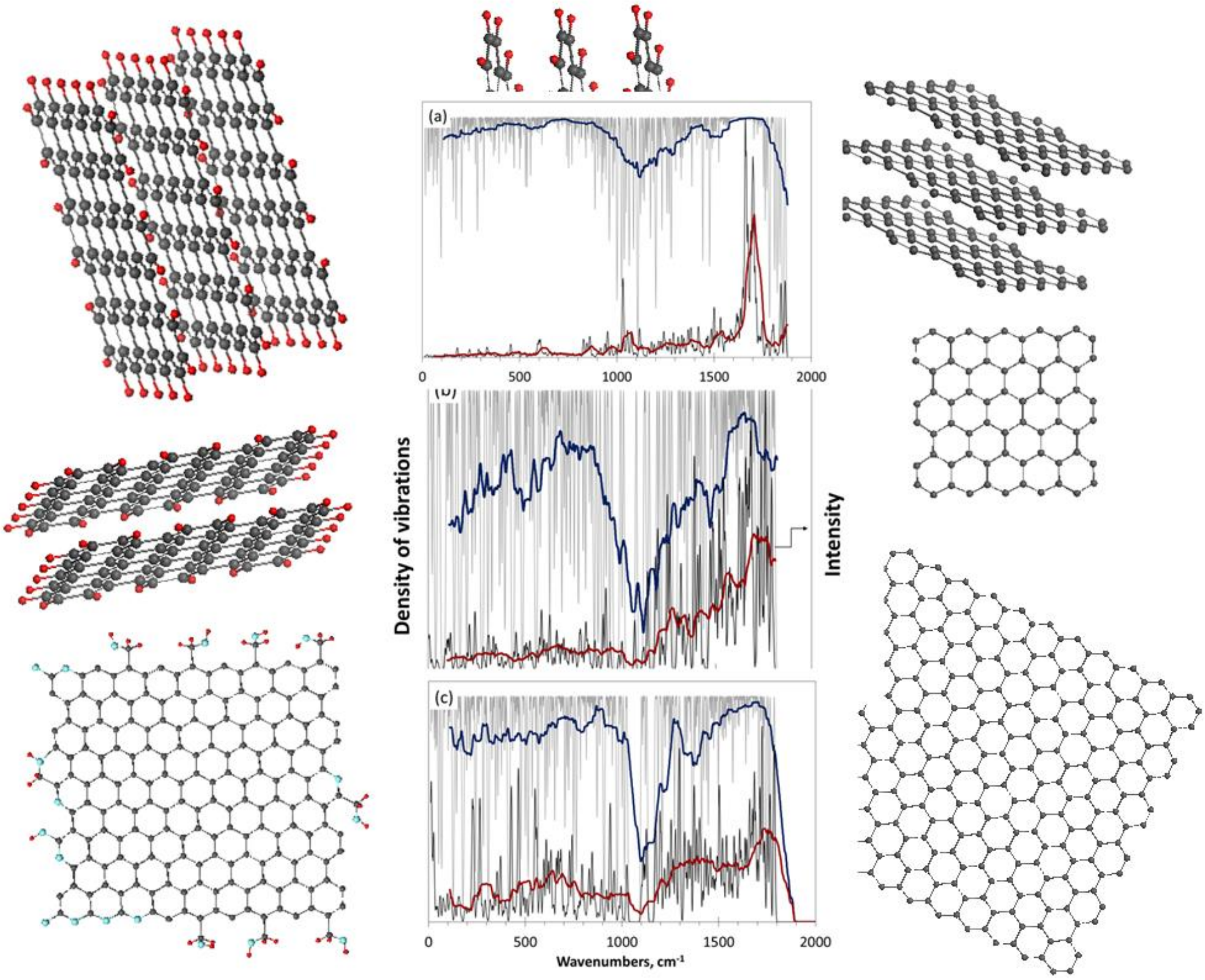

**Figure 1.5**. Raman spectra of DTs with different size of graphene domains and of different domain packing. (a, b). One-layer TE-rGO BSU based on domain (9,9) NGr ($C_{192}O_{19}H_{44}$) and bare domain (5,5) NGr ($C_{66}$), respectively; (c). Two layers of hydrogenated domain (5, 5) NGr ($C_{132}H_{44}$). Raman spectra and density of vibrations are accompanied with 50-points trend lines. UHF AM1 calculations.

### *GO in light of the Digital Twins concept*

*Digital twins of the first approach.* Digital twins GO1 and GO2 were synthesized attempting to realize in their structure all the information about the chemical composition of real GO available by that time [27]. The domain (5,5) NGr was taken as the basis for the synthesis, the first stage of which, involving the domain edge atoms, was accompanied by a successive consideration of the addition of single oxygen atoms, hydroxyls, and carboxyls at each step. After evaluating the binding energy (BE) of each attachment separately, the choice of the obtained graphene domain derivatives was made in favor of the configuration with the highest BE. For this derivative, the highest-rank ACS points the number of a target atom of the next oxidation step. This process continued with 22 steps, as a result of which the necklaced graphene molecule $C_{66}O_{22}$ was formed, and the oxidation moved to the basal plane. Each step of this process was considered as a choice between the involvement of a $sp^2$C-C bond in the formation of either a $C_2O$ epoxy group or opening the bond with the addition of two hydroxyl groups thus forming $C_2(OH)_2$ composition. In addition, the landing of heteroatoms *up* or *down* in the case of GO1 and only *up* in the case of GO2 was controlled.

Virtual vibrational spectra of these DTs are shown in Figure 1.6. The subsequent description of these virtual spectra is based on the general frequency kits (GFKs) and is given in [58]. Looking at the Raman spectra of DTs GO1 and GO2 in Figure 1.6b, one immediately notices a triplet of intense bands at ~1760 $cm^{-1}$ (**I**), ~1900 $cm^{-1}$(**II**), and ~2100 $cm^{-1}$ (**III**). According to the tabulated GFKs, the bands should be attributed to the $\nu\, sp^3$**C-C**, $\nu\, sp^2$**C-C**, and $\nu\, sp^3$**C**=O stretchings, respectively. The latter two assignings are non-alternative while the first one is ambiguous. The $\nu\, sp^3$**C =O** origin of band **III** makes it possible to estimate the unavoidable shift of virtual frequencies in this region with respect to the experimental ones [74], which is a blue one and constitutes 200-300 $cm^{-1}$. As seen in the figure, applying this shift to experimental spectra of a real sample (see a detailed description of the product and its spectra in [75]) leads to an obvious agreement between the calculated and experimental spectra concerning both IR absorption and Raman scattering. In the latter case, the characteristic empirical doublet of D-G bands conveniently covers virtual bands **I** and **II**, making it possible to consider virtual **I-II** bands as wished DTs of the empirical D-G ones. Naturally, this is not about exact replicas, but about agreement in the main elements of the structure interpretation.

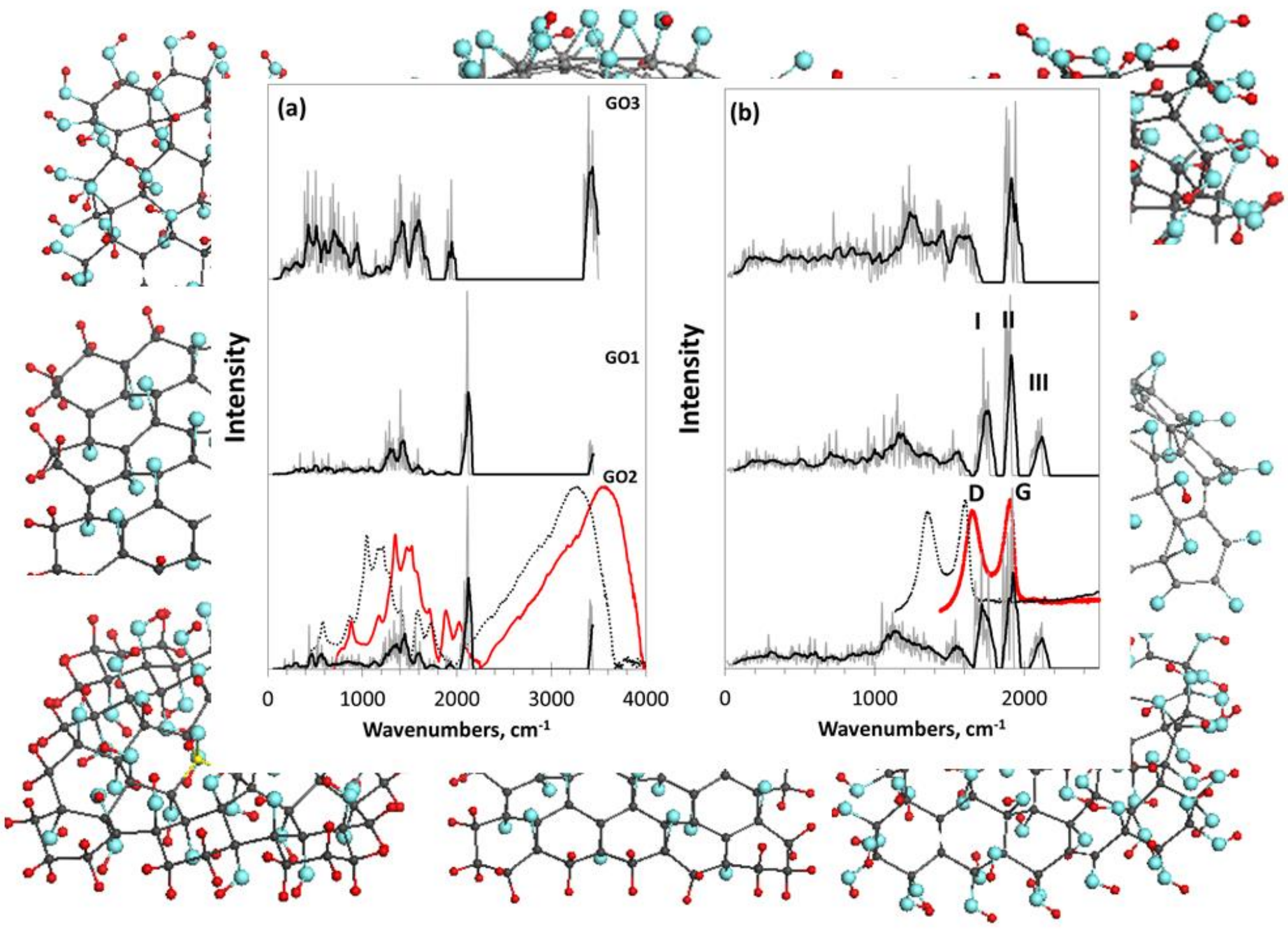

**Figure 1.6**. Virtual one-phonon IR absorption (a) and Raman scattering (b) spectra of digital twins GO1, GO2, and GO3. Spectra plottings are accompanied with trend lines, corresponding to 50-point linear filtration. RHF AM1 calculations. Dotted and red plottings present original and blue shifted on 300 $cm^{-1}$ experimental spectra of GO produced by AkKo Lab Company [75], respectively.

Despite the apparent success in obtaining a doublet of **I-II** bands in the virtual Raman spectra of GO1 and GO2, the main question of the identity of the **I-II** and D-G doublets still remains. As occurred, a design of a particular set of DTs made it possible to obtain the desired answer. Concerning chemical content and linear dimensions, GO1 and GO2 are well consistent with experimental data, which should be attributed to the basic structural units of real GO samples. Unexpectedly, their virtual Raman spectra presented in Figure 1.6b confirmed the validity of these models to even greater extent. The fact is highly exciting but reasons of such an exclusive behavior stays in the shadow. At the same time, the described oxygen status of DTs GO1 and GO2 is quite complicated, too polychrome, and does not allow making any conclusions about reasons of these features beyond speculations. Evidently, the monochromization of the oxide composition may be a logical step towards overcoming this polychrome difficulty and solving the spectral riddle. The first attempt was made when replacing complex DTs GO1 and GO2 with GO3. Virtual synthesis of the latter, based on the same domain (5, 5) NGr, was performed with the participation of hydroxyls only [27]. As can be seen from Figure 1.6b, the monochromization drastically affects both IR and Raman spectra. As for the latter, only the band **II** is clearly observed in the Raman spectrum of the molecule, the position of which coincides with that one in the GO1 and GO2s spectra, while the band **I** is drastically smoothed. Understanding that the presence of a full **I-II** doublet in the Raman spectra of GO1 and GO2 is provided with their oxygen polychrome content, results in the necessity of a particular 'decomposition' of the latter to isolate that component of this polychromicity, which determines the existence of both bands **I** and **II**. It turned out to be possible to implement such a decomposition using a set of DTs specially configured.

*Digital twins of the second approach*. The relevant DTs were based on the domain (5,5) as well while differing with either monochrome necklace content around edge atoms (carbonyls and methylene ones) or filling the basal plane with either epoxy or hydroxyl groups [58]. IR and Raman spectra of these DTs respond to this action differently. Concerning IR absorption, the replacement of carbonyl necklace with methylene one results in a drastic changing of the spectra. At the same time, the spectra of DTs with the same necklaces evidently conserved a common character, although quite markedly influenced by events occurring in the basal plane of the domain. As for Raman spectra, none of monochrome necklaces is responsible for the **I-II** band doublet. The same relates to the hydroxyl covering of the domain basal plane while the epoxy covering directly influences this region structure, although reproducing the doublet band **I** only. The situation is similar to that previously discussed for the DTs spectra in Figure 1.6b, which allows suggesting that the band appearance is tightly connected with the epoxy coverage of the domain basal plane. Since the carbon carcass of the current oxides is formed with the $sp^3$C-C valence bonds only, the band position is associated with the formation of specific $sp^3$**C-C** stretchings strongly affected by the presence of oxygen atoms. A peculiar behavior of vibrational bands in organic molecules associated with oxygen atoms was noticed as early as 40 years ago [76, 77] and since then has been constantly discussed in the literature (see [78-81] and references therein). These bands easily changed their position and intensity depending on the place of these atoms in the molecule, on their number, on the presence of certain heteroatoms, and so forth. However, until now it does not go beyond the statement of the fact as well as attempts to link the latter with the presence of special valence bonds. In the case of the discussed DTs, which are oxygen-polytarget objects, nature itself provides us with a unique way to relate the observed spectral effect to the structure of valence bonds. Performing the virtual synthesis of polyderivatives of the graphene domain gradually following the ACS spin-density algorithm discussed earlier, one can fix the structural composition of covalent C-C bonds characteristic of each step of the reaction.

Thus, DT GO4, fully covered with oxygen atoms, attached to the domain edge atoms in the form of carbonyls C=O and coupled with basal-plane atoms forming $C_2O$ epoxy groups, was the best to explain the appearance of the band **I** and to satisfy the large oxygen content of the empirical GOs. However, the reproduction of the D-G-doublet empirical spectrum with virtual **I-II** doublet is still incomplete and requires explanation. As said earlier, the band G position indicates, without alternative, its assignment to $sp^2$**C-C** stretchings, which, in turn, evidences the presence of unoxidized carbon atoms and, consequently, of $sp^2$C-C covalent bonds in the GO carcass of real samples. It means that the $sp^2$-to- $sp^3$ transformation, which accompanies oxidation of bare graphene domains, is incomplete. As said, the ACS-spin-density driven DTs design allows monitoring a gradual $sp^2$-to- $sp^3$ transformation of the covalent bond structure of the GO carbon carcasses. Thus, to perform such a monitoring, DTs of the third approach were required. Those present derivatives of the step-by-step oxidation of the reference $sp^2$ rGO (GO4_00, $C_{66}O_{22}$), which concerns the oxidation of edge atoms of the basic domain (5,5) NGr, up to the completely oxidized $sp^3$ GO4 (GO4_22, $C_{66}O_{44}$). Complete coverage of the 44-atom basal plane of the reference domain with epoxy groups required 22 steps [58].

*Monochrome digital twins of GO of the third approach.* As was shown, the location of the first epoxy group on the basal plane (DT GO4_01) results in a significant reconstruction of at least 20 bonds of the reference $sp^2$C-C bonds pool, which is caused by the expected elongation of the bonds [58]. Confirming that Raman scattering produces a spectral signature of the C-C bond configuration [25], the DT Raman spectrum in comparison with the reference GO4_00 one, reveals a sharp response to the change in the bond structure. This feature accompanies the addition of each next epoxy group. Thus, the second epoxy group addition reveals the bond transformation of DT GO4_02 more markedly and this effect increases at the subsequent steps. Since the observed $sp^2$-to-$sp^3$ transformation of the honeycomb structure is accompanied with unavoidable mechanical stress, the compensation of the latter causes not only expected elongation of the reference $sp^2$C-C bonds, but also its considerable shortening to protect the whole carbon body from destruction.

Figure 1.7 present the final stage of the reaction monitoring. The last steps are followed with severe decreasing of the ACS values, which are zeroing at the 17$^{th}$ step. Simultaneously, the bond length dispersion $\Delta l_{C-C}$ drastically decreases, which causes a remarkable structure ordering of the bonds pool of GO4_17 with respect to preceding ones. The bond distribution reveals two types of bonds, first of which concerns the main massive of single $sp^3$C-C ones of 0.150±0.001 nm in length, while the second is related to five highly shortened double $sp^2$C-C bonds of 0.135±0.0001 nm. Evidently, the appearance of abnormal double bonds results from the need to compensate the accumulated mechanical stress, caused with long single bonds. However, shortening the $sp^2$C-C bonds below $R_{crit}$ = 0.1395 nm leads to zeroing ACS due to drastic dependence of the latter on the corresponding bond length [66]. Therefore, the appearance of shortened $sp^2$C-C bonds means stop further chemical reaction since the corresponding carbon atoms are not more chemically active targets, so that the oxidation of basal-plane atoms retains incomplete.

As seen in Figure 1.7c, a dramatic change in the Raman spectra occurs at the 17th step of the addition of the epoxy group to the basal plane simultaneously with the structural transitions described above. The spectrum becomes much more structured and the manifestation of both bands **I** and **II** is clearly observed. Further oxidation can be performed manually only by random placing epoxy groups over five remaining double $sp^2$C-C bonds. As seen from the figure, the distribution of bonds over lengths of the last-step completely $sp^3$ GO4 (GO4-22), involving a full set of attached epoxy groups, practically does not change in this case. However, the former shortened double $sp^2$C-C bonds become single $sp^3$ ones and the band **II** of the Raman spectrum disappears. Thus, simultaneous monitoring of the per-step synthesis of the oxide by the C–C bonds pool and Raman spectra allowed concluding that the **I**-**II** doublet in the oxide spectrum is due to the establishment of a balanced structure of highly ordered $sp^3$ and $sp^2$ C-C bonds, which keeps the carbon skeleton the most stable and undestroyed. Meeting both requirements is accompanied by an obvious structural ordering of the skeleton along the lengths of the bonds, which manifests itself in a small and almost vanishing dispersion of $sp^3$ and $sp^2$ C-C bonds, respectively.

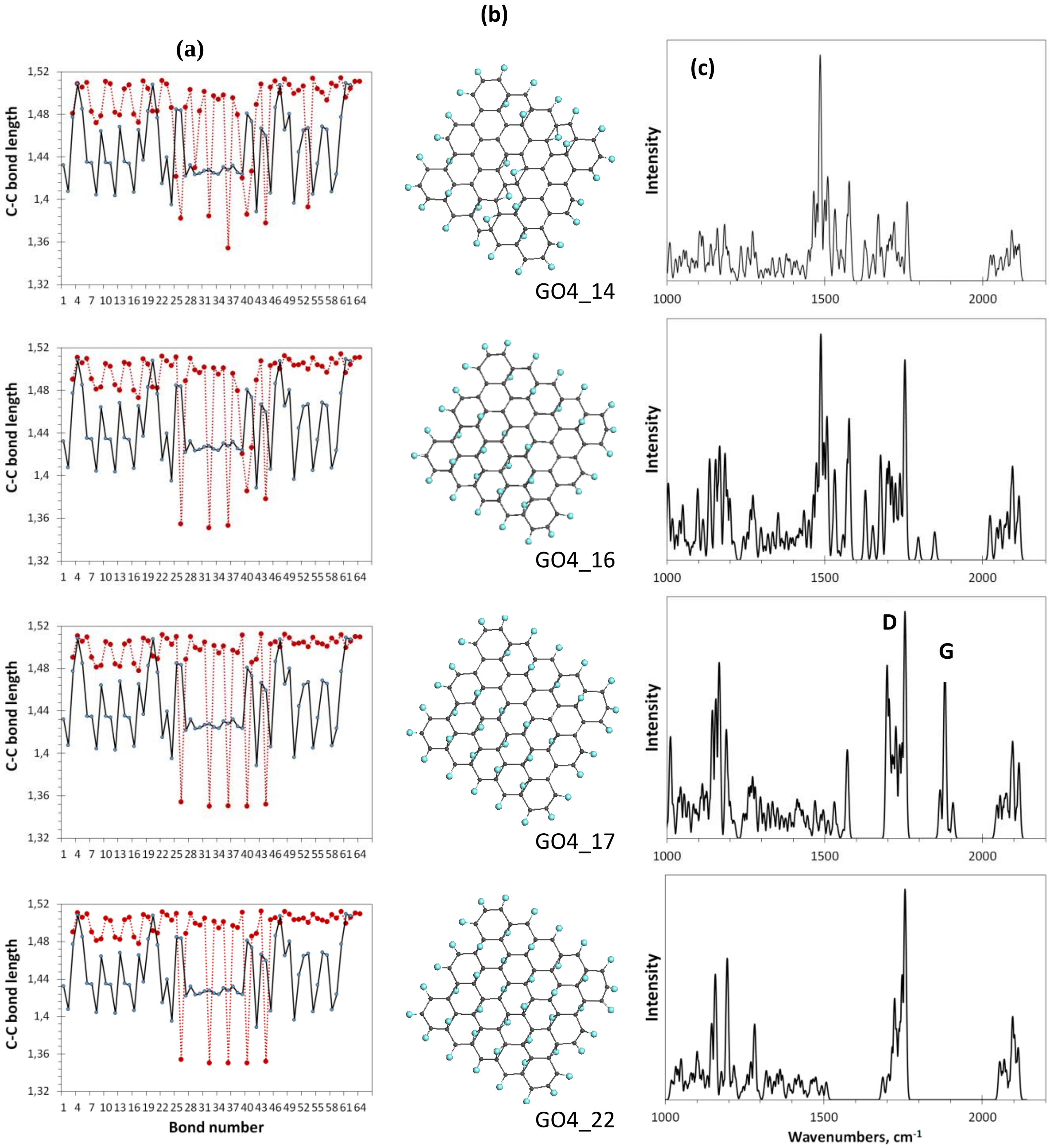

**Figure 1.7**. Monitoring of the $sp^2$-to-$sp^3$ C-C bond structure transformation at final steps of the oxidation of basal-plane carbon atoms of GO4. a. TD C-C bond length (reference and current presented with black and dotted red plottings, respectively). b. TD equilibrium structure. c. TD Raman spectrum. UHF AM1 calculations.

The assignment of bands **I** and **II** in the Raman spectrum of GO4_17 to the corresponding features of their vibrational spectra makes us to return to Figure 1.5b and check, whether the virtually synthesized oxides GO1 and GO3 meet this requirement. As occurred, both DTs correspond to incomplete oxidation, which is terminated on the 18 and 38 steps of the reagent placing on the basal plane of the pristine domain, respectively, due to ACS zeroing [27]. In both cases, this reaction step is accompanied by the formation of a few shortened $sp^2$C–C bonds of 0.135±0.0003 nm in length. It is these low dispersed bonds, which cause the appearance of the band **II** in the virtual spectra of GO1 and GO3, once band G in experimental Raman spectra of the GO.

Hereby, answering the questions posed at the beginning of the section, we can state the following. The standard form of Raman spectra of GOs has a chemical origin. The D and G bands are due to stretching vibrations of $sp^3$C-C and $sp^2$C-C bonds, respectively. In this case, $sp^3$ bonds constitute the main structural array of the GO graphene carcass, while $sp^2$ bonds are the result of stopping the oxidation of basal-plane carbon atoms. Accordingly, the main conclusion concerning the identity of Raman spectra of rGO and GO is the following. The identity does take the place and is due to the $sp^3$ and $sp^2$ character of D and G bands, respectively. These high frequency stretchings correspond to the main pools of the bodies covalent bonds and are manifested as D and G bonds of GO and rGO, respectively. Their $sp^2$ (G) and $sp^3$ (D) companions are of different origin, once presenting high frequency stretchings of unoxidized $sp^2$C-C bonds caused by stopping oxidation of the graphene domain basal plane, thus marking incomplete oxidation of GO as a characteristic feature, and of dynamical $sp^3$C-C bonds resulted from the rGO stack structure. Over 400 DTs, taking together and subjected to the digital device HF Spectrodyn, have allowed clarifying this exclusive spectra case.

### Discussion on Digit Twins concept and conclusive remarks

Concluding this chapter, let's try to answer two more important questions: why the exceptional spectral phenomenon we have considered has not been explained until now, provided that theoretical approaches and modeling in graphenics have acquired such a large scale that they have turned it into an almost virtual science? And why didn't numerous simulations lead to the DT concept that was so successful?

From the author's standpoint, there are two reasons that explain the current situation. Explaining them, we will keep in mind Scheme 2, which describes the essence of the DT concept. So, the first reason concerns the wrongful vision of the object or DT. Theorists consider rGO or GO from the positions of graphene crystal, regardless of the nanosize of their basic structural units and, thus, not taking into account any of size effects. The results obtained in the framework of the theory of the solid were offered in an obsessive way to be applied to real objects. It suffices to mention the ratio of the intensities of the D and G bands of the Raman doublet, $I_D/I_G$ , which is widely used under unremitting pressure, to determine the defectiveness and size of graphene domains [50, 82, 83], thereby attributing to the band D the ***d***efect origin. However, as expected, when these structural parameters were determined independently, no relationship between them and the intensity ratio was found [65]. The second example concerns the identity of the Raman spectra of rGO and GO considered in the chapter. According to the theory [84], this identity is due to the practical identity of the graphene domains of rGO and GO. The existence of more than 50 at% oxygen in the latter case was able to cause a shift of the broad G band in the GO spectrum to the short-wavelength region by ~14cm$^{-1}$. No $sp^2$-to-$sp^3$ transformation of the covalent bonds of the solids was considered. No conclusion was made tp couple the G band existence with the predetermined incomplete oxidation of the graphene domain in the basal plane and the preservation of its inseparable structure. As for band D, as usually, it was considered as the sign of the graphene domain defectiveness.

Quantum chemists tried to keep the vision of the nanoscale object. However, the calculation programs used in the overwhelming majority of cases provided for the use of solid-state algorithms for calculating crystals with periodic boundary conditions, and therefore the proposed models were reduced to unfolded supercells. Both physicists and quantum chemists, in the overwhelming majority, did not take into account the radical nature of the graphene domain and the features associated with it. This led to the second reason that concerns digital device in Scheme 2.

The polyatomic nature of the models and/or the corresponding supercells required the use of fast and efficient calculation techniques. Only two the most powerful semi-empirical methods were considered, those based on the Hartree-Fock (HF) approximation and density functional theory (DFT). It so happened that the development of these methods was accompanied by intense competition, as a result of which the fans of the DFT method won (see a detailed discussion of the problem in [85] and references therein) and virtual graphenics is predominantly DFT one. However, the DFT approach, which is quite effective in the case of closed-shell molecules, turns out to be helpless when considering radicals with open shells. Concerning the problem, the DFT mainstream agreed to forget about the radical properties of graphene, thus opening up a wide road for the DFTzation of virtual graphenics. But the use of an inadequate virtual device in Scheme 2 either inevitably leads to incorrect results or does not allow you to see the essence of the process in question. Thus, the DFT Spectrodyn, if used, does not allow for controlled virtual synthesis of DTs of the required quality and has to be restricted with hand made paintings only. It significantly distorts the consideration of out-of-plane displacements of carbon atoms of neighboring layers domains, and does not allow detailed monitoring of the $sp^2$ -to-$sp^3$ transformation of the covalent C-C bonds. As a result, the mystery of the identity of the Raman spectra of rGO and GO for the DFT device remains closed forever.

This misfortune concerns not only the considered case, but also all the DFT results, following which obscured physical and chemical meaning of the objects under study led to the sad results of the extended Graphene Flagship program. In contrast, a conscious and controlled design of DTs, an adequate virtual device and a wide comparative analysis of the obtained IT results, indeed, make it possible to sort out intricate problems, to reveal such hidden features as the ban on 100% oxidation of a simple graphene domain, provided in advance by nature.

The proposed analysis of the vibrational spectra of rGO and GO is based on the important concept of the radical nature of bare graphene domains and rGO. Many spears have been broken in discussions that cast doubt on the radical nature of graphene materials. The author would like to hope that the above described study will allow rejecting the last remaining doubts and will force researchers to work with materials familiar to them in a new way. This conclusion directly concerns the difference of carbon and silicon natural species.

Both elements are adjacent in the column of the periodic table, do not exist in the liquid phase, exhibit a number of similar properties in the gaseous and solid state, but at the same time they have a number of significant differences. Limiting ourselves to a solid body, we note such important points as the absence of a graphite-like structure in silicon, in contrast to the wide distribution of graphite, the relatively easy accessibility of the cubic form of solid silicon, in contrast to the extremely hard one in the case of diamond, and the widespread occurrence of silica in nature, in contrast to the complete absence of carbon oxide. These and other features of the silicon-carbon pair have been the subject of discussion since ancient times. However, the explanation of possible causes began to form relatively recently, which coincided with the emergence of graphenics, which was the stimulus for a wide discussion of this problem. It all started with the phantom of silicene and ended with the understanding that the reason for the discussed differences is the prohibition of the implementation of $sp^2$ hybridization for silicon atoms, in contrast to carbon ones [14]. Silicon is the $sp^3$ element only, while solid carbon can exist in all three forms of hybridization: $sp^1$(see about solid acetylene [86]), $sp^2$, and $sp^3$. It was found that a significant increase in the size of silicon atoms and, accordingly, the Si-Si interatomic covalent distances prevents the realization of $sp^1$ and $sp^2$configurations of atoms due to the exceptionally high radicalization of the corresponding atomic compositions, leaving the possibility of stable existence only for the $sp^3$ forms.

The basic structural unit of silica is the silicon-oxygen tetrahedron $SiO_4$, whose structure ideally reproduces the $sp^3$ hybridization of silicon atoms. In contrast, the carbon-oxygen tetrahedron $CO_4$ is extremely unstable, and even in those cases where its formation is implied, the carbon atom is in $sp^2$ hybridization [87]. This circumstance, as well as the obviously very severe conditions for the formation of diamond (40 GP and 960-2000$^0$C) [89], are natural and rule out the possibility of the formation of carbon "silica" or diamond oxide. Unlike $sp^3$ diamond, $sp^2$graphite is widely distributed; however, graphite oxide does not exist in nature. The reason for this has recently been considered in relation to another natural carbon material – shungite carbon [40]. It was suggested that the formation of an extended $sp^2$ material determines the rate of growth of primary carbon lamellae. The graphene lamella grows until its edge carbon atoms are completely terminated by surrounding heteroatoms, the most active of which are oxygen ones. The freezing of growth and the continued presence of such a lamella in an oxygen-containing atmosphere is naturally accompanied by the oxidation of atoms lying in the basal plane, accompanied by the $sp^2$-to-$sp^3$ transformation. However, the presence of water as a reducing agent [89], which usually accompanies graphite deposits, leads to the restoration of the previous $sp^2$ configuration caused by the removal of oxygen atoms from the basal plane. During a long time of graphite formation, the processes of oxidation and reduction settle down and the latter wins as more thermodynamically favorable. Graphite oxide, which is absent in nature, is obtained only synthetically by hard oxidation of crushed graphite during a regulated technological cycle [30-38].

Such a long conclusion was needed to draw the reader's attention, firstly, to how important the permissibility of the $sp^2$-to-$sp^3$ transformation is in the life of solid carbon, and, secondly, to the fact that graphene oxide and reduced graphene oxide are a close pair tightly connected by this transformation to each other. In this shapter, we showed that it is this unique property of the structural rearrangement of carbon atoms that allows answering the question of the reasons for the identity of the Raman spectra of the parental and reduced graphene oxides.